\documentclass[12pt, a4paper]{article}
 \usepackage[T2A]{fontenc}
 \usepackage[cp1251]{inputenc}
 \usepackage[russian, english]{babel}
 \inputencoding{cp1251}
 \usepackage{latexsym}
 \usepackage{amssymb}
 \usepackage{bibunits}
 \usepackage{wrapfig}\usepackage{ifpdf}
 \ifpdf
         \usepackage[pdftex]{graphicx}
         \usepackage{epstopdf}   
 \else
         \usepackage[dvips]{graphicx}
 \fi
 \graphicspath{{./images/}}
 \usepackage{epsfig,color}
 \usepackage{float}
 \usepackage{caption}
 \usepackage{subcaption}
 \graphicspath{{images/}}
 \usepackage{titlesec}

\titleformat*{\section}{\normalsize\bfseries}

\begin{document}

%{%
%\selectlanguage{english}

%\udc{531.395:512.622}

\title{ \Large\bf 
Algebraic dynamics on a single worldline: \\ 
Vieta formulas and conservation laws} 
\author{\normalsize  V. V. Kassandrov, I. Sh. Khasanov and N. V. Markova} 
\date{\small \textit{Institute of Gravitation and Cosmology, \\ Peoples' Friendship University 
of Russia, Moscow, Russia} \\ 
%\tt e-mail: vkassan@rambler.ru khasanov@sci.pfu.edu.ru
}

\maketitle

\begin{abstract}

In development of the old conjecture of Stuckelberg, Wheeler and Feynman 
on the so-called "one electron Universe", we elaborate a purely
algebraic construction of an ensemble of identical pointlike particles occupying the same worldline and 
moving in concordance with each other. In the proposed construction one does
 not make use of any differential equations of motion, Lagrangians, etc. Instead, 
we define a ``unique'' worldline implicitly, by a system of nonlinear
  polynomial equations containing a time-like parameter.  Then, at each
  instant, there is a whole set of solutions defining the coordinates of particles-copies localized on the unique worldline and moving along
  it. There naturally arise two different kinds of such particles which correspond to 
real or complex conjugate roots of the initial system of polynomial equations, respectively. At some particular instants, 
one encounters the transitions between these two kinds of particles-roots that 
model the processes of annihilation or creation of a pair ``particle-antiparticle''.   We restrict by consideration of nonrelativistic collective dynamics of the ensemble of such particles on a plane. Making use of
  the techniques of resultants of polynomials, the generating system
  reduces to a pair of polynomial equations for one unknown, with
  coefficients depending on time. Then the well-known Vieta formulas
  predetermine the existence of time-independent constraints on the
  positions of particles-roots and their time derivatives. We
  demonstrate that for a very wide class of the initial polynomials
  (with polynomial dependence of the coefficients on time) these
  constraints always take place and have the form of 
  the conservation laws for total momentum, angular momentum and (the
  analogue of) total mechanical energy of the ``closed'' system of
  particles.

\end{abstract}

\section{Introduction}

In the presented paper we elaborate an algebraic realization of the
ideas of Stueckelberg, Wheeler and Feynman on the possibility of
unified dynamics of an ensemble of identical pointlike particles
located at different places of a {\it unique
  worldline}. E.C.G. Stueckelberg~\cite{Stueckel1, Stueckel2} was,
perhaps, the first who considered a worldline containing segments with
superluminar velocities and forbidden, therefore, in the canonical
theory of relativity; this assumption explicitly results in the notion
of a ``multi-particle'' worldline.  Later on, J.A. Wheeler (in his
famous telephone call to R. Feynman, see~\cite{FeynmanNobel})
suggested his concept of the ``one-electron Universe'' that allows for
the above mentioned construction of an ensemble of numerous copies of
a single particle that all belong to the same worldline.  One of the
consequences of these ideas, namely, the consideration of a positron
as a ``moving backwards in time'' electron, had been later exploited
by R. Feynman in his construction of quantum
electrodynamics~\cite{FeynmanPositrons}.  However, the very
conjecture of the ``one electron Universe'', for many reasons, had
been abandoned for a long time.
 
In recent paper~\cite{Khasanov}, in the framework of a purely
non-relativistic, Newtonian-like scheme, we have made an attempt to
deduce the correlations in positions and movements of different
copies-particles from algebraic properties of their (common)
worldline, without any resort to equations of motion (Newton's laws,
Hamiltonians, Lagrangians, etc.) themselves.  Specifically, we
considered the (unique) ``worldline'' which, instead of the generally
accepted {\it parametric} form $x_a = f_a (t)$, $a=1,2,3$, is defined
{\it implicitly}, by a system of algebraic equations
\begin{equation}\label{implicit}
F_a (x_1,x_2,x_3,t) =0.
\end{equation}

\begin{figure}[ht]
\includegraphics[width=80mm]{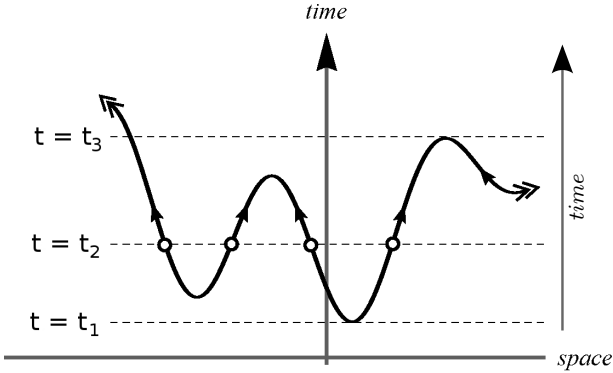}
\caption{Generic worldline, numerous pointlike ``particles'' (at $t=t_2$) and   
creation (at $t=t_1$) or annihilation (at $t=t_3$) events}
\label{img:figure1} 
\end{figure}

Then, for any value of the time-like parameter $t$, one generally 
has a whole set $(N)$ of solutions to this system, which define a 
correlated kinematics $x_a = f^{(k)}_a (t),~~k=1,2,\ldots N$ of an ensemble of 
identical pointlike singularities on a unique worldline (fig.\ref{img:figure1}). 

If, in the course of time, the parameter $t$ is assumed to increase
monotonically, then a pair of particles can appear at a particular
instant $t=t_1$ (or disappear at $t=t_3$). These ``events'' model the
processes of creation (annihilation) of a pair
``particle-antiparticle''.

In the paper~\cite{Khasanov}, we restricted our consideration to the plane (2D) motions and a
\textit{polynomial} form of the generating functions $\{F_a\}$ in (\ref{implicit}). The latter restriction allows for a
complete determination of the full set of roots-particles of the
system and explicitly reveals the correlations in their positions and
dynamics represented by the well-known {\it Vieta formulas}. Moreover,
in a special ``inertial-like'' reference frame, the first (linear in
roots) Vieta formula implies the {\it
  conservation law for total momentum} of the set of pointlike
particles identified with the roots of (\ref{implicit}).  In this way
one reproduces the general structure of Newtonian mechanics which,
thus, turns out to be {\it encoded in the algebraic properties of an
  arbitrary (implicitly defined) worldline}! The most important
results and ideas of the paper~\cite{Khasanov} are presented in
section 2.

Apart of the obvious program of relativistic invariant formulation of
the scheme (for this, see discussion in~\cite{Khasanov}, section 5),
there is a number of problems which can be successfully considered in
the framework of purely Galilean-Newtonian picture. One of these is
the determination of the full set of {\it conservation laws} which
follow from the structure of the {\it nonlinear} (i.e., of higher
degree in roots) Vieta formulas. These issues are examined in section
3. The problem of conservation of total {\it angular momentum} is
separately treated in the next section 4 and turns out to relate to
the Vieta's constraints as well. In this section we also present 
a typical example of the polynomial system that defines a self-consistent
dynamics of a number of roots-particles subject to all three
canonical-like conservation laws.  Section 5 contains some concluding
remarks and discussion.

\section{Collective Algebraic Dynamics on an Implicitly Defined
  Worldline}

\begin{figure}[ht]
\includegraphics[width=0.5\textwidth]{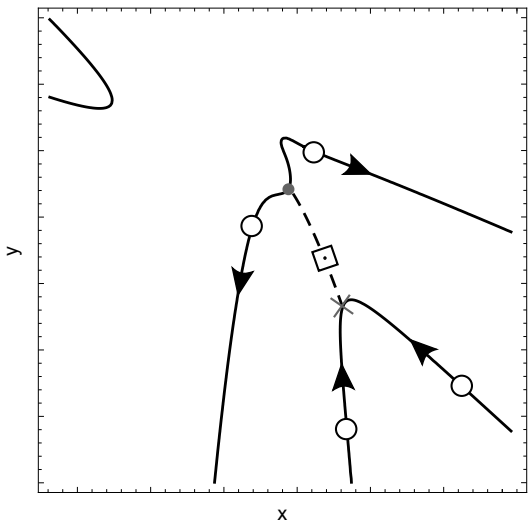}
\caption{Three branches of the trajectory of R-particles and the
  typical succession of events (annihilation -- propagation of a
  C-particle --- creation)}
\label{img:figure2}
\end{figure}

Consider, for example~\cite{Khasanov}, a simple algebraic system of
polynomial equations selected quite randomly (yet of rather low
degrees in the unknowns $x,y$ and the time-like parameter $t$):
\begin{equation}\label{example} \left \{ 
\begin{array}{ll}
 F_1(x,y,t) = -2 x^3+y^3+tx+ ty+y+2 = 0, \\ 
 F_2(x,y,t) = -x^3-2 x^2 y+t+3 = 0.
\end{array}
\right. 
\end{equation}                           
Eliminating $t$, one obtains the trajectory 
\begin{equation}\label{trajex}
x^4+3x^3y+2x^2y^2-2x^3+y^3-3x-2y+2 =0  
\end{equation}
which consists of {\it three disconnected components}
(fig.~\ref{img:figure2}).  Solving now the second equation with respect to $y$ 
and substituting the result in the first one, we obtain a 1D polynomial equation of the form 
\begin{equation}
\label{exampleX} \left\{
\begin{array}{ll}
P(x,t) = -17x^9+(4t-4)x^7+(3t+25)x^6 +(4t^2+16t+12)x^4 +\\
+(-3t^2-18t-27)x^3 +t^3+9t^2+27t+27=0.
\end{array}
 \right.
\end{equation}
It can then be proved (say, via the resultants' techniques, see section 3) that there is 
an analogous condition for the unknown $y$, namely,  
\begin{equation}
\label{exampleY} \left\{
\begin{array}{llll}
Q(y,t) = 17y^9+(33t+35)y^7+(-6t+52)y^6+(15t^2+34t+19)y^5 +\\
+(-16t^2+8t+40)y^4 +(-t^3+11t^2+49t+113)y^3+\\
+(-18t^3-72t^2-50t-12)y^2 +(28t^3+148t^2+208t+48)y+ \\
+t^4-48t^2-5t^3-96t-64=0.
\end{array}\right.
\end{equation}

For any value of the time-like parameter $t$ equation (\ref{exampleX})
has nine solutions $x_k,~k=1,\ldots, 9$, each one of which is in
correspondence with a solution $y_k$ of (\ref{exampleY}) so that the
pairs compose exactly nine solutions of the system (\ref{example}).
Some of these are {\it real-valued}, correspond to the so-called {\it
  R-particles} and belong to one of the three branches of the
trajectory (\ref{trajex}) while others arise as {\it complex conjugate
  pairs} and should be treated as another kind of {\it composite}
particle-like formations, the so-called {\it
  C-particles}~\cite{Khasanov}.  The latter can be visualized via
equal real parts of complex conjugate roots and are located in the
space off the trajectory~(\ref{trajex}),~see fig.\ref{img:figure2}.

At particular values of the time-like parameter $t$ defined by the
condition (we rename here $x_1=x,~x_2=y$)
\begin{equation}\label{multiple}
\det \left\Vert \frac{\partial F_A}{\partial x_B} \right\Vert = 0,\quad A,B,\ldots =1,2, 
\end{equation}
some two real roots of the system (\ref{example}) merge together and
transform then into a complex conjugate pair or vice versa. These {\it
  events} can be evidently interpreted as the processes of {\it
  annihilation} of a pair particle-antiparticle accompanied by the
formation of a ``{\it C-quantum}'', or {\it creation} of a pair,
respectively. A typical succession of processes is represented at
fig.~\ref{img:figure2}.  Note that, in contrast to the picture earlier
suggested by Stueckelberg, all these processes necessarily satisfy a
number of conservation laws, see below.

Consider now the set of {\it Vieta formulas} specific, say, for the
polynomial equations (\ref{exampleX},\ref{exampleY}) and explicitly
representing the correlations in positions and dynamics of different
particles-roots defined by the generating system (\ref{example}). The
first two of them (of lowest degrees in roots) have the following
form:
\begin{equation}
\label{linear}
x_1+x_2+\ldots + x_9 = 0, \quad y_1+y_2+\ldots +y_9=0;
\end{equation}
and, respectively, 
\begin{equation}\label{quadratic}
x_1x_2+x_1x_3+\ldots +x_8x_9=-\frac{1}{17}(4t-4), ~~\\
y_1y_2+y_1y_3+\ldots +y_8y_9= \frac{1}{17}(33t+35),  
\end{equation}
where all the roots depend on $t$ and the zeros on the right-hand side of 
equations (\ref{linear}) are direct consequences of the 
absence of the 8-th degree terms in $P(x,t)$ and $Q(x,t)$, see 
(\ref{exampleX},\ref{exampleY}). Recall now that  
for a mechanical system of nine particles with masses ${m_k}$ the coordinates 
of its center of mass are 
\begin{equation}\label{centerXY}
X(t)=\frac{1}{9}(m_1x_1+\ldots +m_9x_9),\quad
Y(t)=\frac{1}{9}(m_1y_1+\ldots +m_9y_9). 
\end{equation}
From the {\it linear Vieta formulas} (\ref{linear}) we conclude,
therefore, that for the polynomial system (\ref{example}) that defines
the ensemble of nine particles {\it of equal masses}~\footnote{Note,
  however, that the masses of {\it composite} {\it C}-particles should
  be, of course, considered twice greater than those of real {\it
    R}-particles}, {\it the center of mass is at rest}. Repeatedly
differentiating then equations (\ref{linear}) with respect to the
time-like parameter $t$, one obtains for instantaneous velocities
$\{\vec v_k(t)\}$ and accelerations $\{\vec a_k(t)\}$ of the
roots-particles:
\begin{equation}
\label{momentum} 
{\vec v}_1+{\vec v}_2+\ldots+{\vec
    v}_9=0,
\end{equation}
and 
\begin{equation}
\label{accelerat}
{\vec a}_1+{\vec a}_2+\ldots +{\vec a}_9 =0.
\end{equation}
Equation (\ref{momentum}) demonstrates that the {\it total momentum of
  nine particles of the ensemble is conserved} and, precisely, equal
to zero.  As for the equation (\ref{accelerat}), one can {\it
  identify} the acceleration $\vec a_k$ with the {\it resulting force}
$\vec F_k$ acting on the particle (of unit mass, $m_k=m=1$) from all
other particles of the ensemble~\footnote{The problem of decomposition
  of ``resulting forces'' into partial forces of mutual interaction is
  rather difficult and will be discussed below; see, e.g., p.10}. Now one can say
that equation (\ref{accelerat}) represents in fact the ``weakened''
form of the Newton's third law: {\it the sum of all resulting forces
  acting on all the nine particles in the ensemble is identically
  zero}. Note that the velocities (momenta), accelerations and other
physical quantities of {\it C}-particles can be always considered
real-valued since they correspond to complex conjugate pairs of roots,
so that their imaginary parts cancel and do not enter the constraints
(\ref{momentum},\ref{accelerat}).
 
Let us examine now the next, quadratic in roots, Vieta formulas
(\ref{quadratic}).  It is easy to see that, making use of the linear
formulas (\ref{linear}), the former can be equivalently rewritten for
the {\it sums of squares of the roots} as follows:
\begin{equation}
\label{squares}
x_1^2+x_2^2+\ldots +x_9^2=\frac{2}{17}(4t-4),\quad 
y_1^2+y_2^2+\ldots +y_9^2=-\frac{2}{17}(33t+35). 
\end{equation}
Composing now a $SO(2)$-invariant combination of the two equations (\ref{squares}) 
(which only is coordinate-independent and thus physically meaningful), one 
obtains:
\begin{equation}\label{expand}
r_1^2+r_2^2+\ldots +r_9^2 = -(58/17)t - (78/17), 
\end{equation}
where $r_k^2:=x_k^2+y_k^2$. Note that the left-hand side of the equation (\ref{expand}) is real but 
 not positive definite since some roots are complex conjugate. The obtained formula (\ref{expand}), though interesting,
is in fact accidental: we shall see below (section 3) that, generally,
the dependence on the time-like parameter $t$ in the right-hand side
of equations like (\ref{expand}) is rather quadratic than linear.  That is
why, to obtain a physically valuable conservation law, one has to
differentiate it {\it twice}. It follows then:
\begin{equation}
\label{energy}
(v_1^2+v_2^2+\ldots +v_9^2)+({\vec a}_1{\vec r}_1+{\vec a}_2{\vec
  r}_2+\ldots +
{\vec a}_9{\vec r}_9) = const = 0. 
\end{equation}

The first group of terms reproduces the (doubled) total kinetic energy
whereas the second group stands for the analogue of potential energy
(or, more closely, total amount of work of all the resulting
forces). Thus, we obtain a coordinate-free constraint which strongly
resembles the {\it law of mechanical energy conservation}~\footnote{The constraints like (\ref{energy}) 
are also closely related to the {\it virial theorem}, see p.10} but
obviously has a different form!  Nonetheless, the correlated dynamics
of particles defined by the systems like (\ref{example}) is governed not only by the
law of conservation of total momentum but by the energy-like
conservation law (\ref{energy}) as well!

It is also noteworthy that the equality
\begin{eqnarray}\label{ident}
\frac{d}{dt} (\frac{1}{2}v_k^2)\equiv \vec a_k\vec v_k \equiv \vec F_k \vec v_k
\end{eqnarray}
is in fact the well-known {\it balance equation} for kinetic energy
which does not depend on the particular form of interaction and is
valid {\it identically} under the identification of acceleration $\vec
a_k$ with the resulting force $\vec F_k$.

Let us further examine the situation with {\it angular momentum} vector  
\begin{equation}\label{angular}
\vec M = \sum  \vec M_k, \quad M_k:= \vec r_k \times \vec v_k, 
\end{equation}
which for the considered case of a plane motion has only one nontrivial 
component. The procedure of analytic determination of the latter will be 
described in section 4 while at the moment we only announce that for a very 
wide class of generating polynomial systems like (\ref{example}) 
{\it the total angular momentum is also conserved}. In particular, for the 
considered system of equations (\ref{example}) it is precisely zero! Of course, 
this result can be confirmed by direct numerical calculations. 

We can now conclude that the (randomly chosen!) system (\ref{example}), without 
any appeal to the Newton's laws or some other differential equations of motion, 
completely defines  a self-consistent (algebraic in nature) dynamics of 
the ensemble of (two kinds of) particles. This dynamics is subject to a number 
of conservation laws (reproducing, in  most part, the canonical set of these), 
includes mutual transmutations of particles, etc. The detailed analysis and 
full {\it animation} of such, unexpectedly rich dynamics (in which, however, one had not 
considered the angular momentum and the energy-like conservation laws) 
was presented in the previous paper~\cite{Khasanov}. 

Quite naturally, a number of remarkable problems arise at the moment: to what 
extent is the situation represented by the particular system (\ref{example}) 
typical? How wide is the class of polynomial (or even general algebraic) 
systems of equations which ensure the validity of conservation laws? Does there 
exist an {\it exceptional} algebraic system (or a class of these) that 
completely reproduces the Newtonian dynamics, and to what kind of particle 
interaction it corresponds then? What changes will occur during the 3D generalization of 
the presented scheme? In which way one can adapt the construction to the 
requirements of Special Relativity? At least first two of these problems will 
be examined and partially solved below. 

\section{Vieta Formulas Generate Conservation Laws}  

Consider the  general 2D situation when a system of two 
polynomial equations
\begin{equation}\label{polynom} \left \{ 
\begin{array}{llll}  
F_1(x,y,t) = [a_{n,0}(t) x^n + a_{n-1,1}(t) x^{n-1}y+...+a_{0,n}(t) y^n]+...\\
~~~~~~~~~~~~~~~~~~ + a_{0,0} (t)=0,\\ 
F_2(x,y,t)= [b_{m,0} (t) x^m +b_{m-1,1} (t) x^{m-1}y+... +b_{0,m}(t)y^m]+...\\
~~~~~~~~~~~~~~~~~~ + b_{0,0}(t) =0
\end{array} \right.
\end{equation}
is under investigation. Note that only forms of the highest ($n$ and
$m$, respectively) and the least orders are written out in
(\ref{polynom}). The coefficients $\{a_{i,j}(t),b_{i,j}(t)\}$ depend
on the evolution time-like parameter $t$ and take values in the field
of real numbers $\mathbb R$.

All the roots $\{x_0(t),y_0(t)\}$ of such a system are either real or
entering both in complex conjugate pairs. In order to find all the
roots of the system (\ref{polynom}), one must somehow eliminate one of
the unknowns, say $x$, reduce the system to a 1D polynomial equation
in $y$, with coefficients depending on $t$, and make then use of the
{\it fundamental theorem of algebra}. The procedure can be accomplished by 
making use of the
so-called {\it method of resultants} (see,
e.g.,~\cite{Gelfand,Morozov} and our paper~\cite{Khasanov}).  As a
rule~\footnote{In more detail the situation is described
  in~\cite{Khasanov}}, the resulting polynomial will be of degree
$N=nm$, and the sought for equation has the form
\begin{equation}
\label{resultY}    
R_x(y) = g_N (t) y^N + g_{N-1} (t) y^{N-1} + \ldots  +g_0 (t) =0.
\end{equation}
The structure of the resultant (which in this case is often called
{\it eliminant}) is represented by the determinant of the {\it
  Sylvester matrix} (see, e.g.,~\cite{Morozov,Prasolov}).
The coefficients $\{g_{I}(t)\}$ depend on $\{a_{i,j}(t),b_{i,j}(t)\}$.
One can exchange the coordinates, and after the elimination of $y$
arrive at the dual condition
\begin{equation}\label{resultX}    
R_y(x) = f_N (t) x^N + f_{N-1}(t) x^{N-1} + \ldots  +f_0 (t)=0 .
\end{equation}
Note that, generally, the degrees of $R_x(y)$ and $R_y(x)$ eliminants
are the same and equal to $N=nm$ (see, for details,~\cite{Khasanov},
section 3).  Of course, in the case of a rather great $N$, {\it
  algebraic computer programs} should be used to find
the explicit form of eliminants. Then, for any value of $t$, all the
$N$ solutions of (\ref{resultY}) and (\ref{resultX}) over $\mathbb C$
can be (numerically) evaluated and put in correspondence to each other
to obtain $N$ solutions $\{x_k(t),y_k(t)\}, k=1,2,..N$ of the initial
system (\ref{polynom}).

Since any system of two polynomial equations can be reduced to a pair
of dual equations (\ref{resultY}) and (\ref{resultX}) for {\it
  eliminants}, polynomials in one variable each, the well-known {\it
  Vieta formulas} are in fact applicable in the 2D case under
consideration. Specifically, one has the following system connecting
all the roots of (\ref{resultX}) or (\ref{resultY}):
\begin{equation} \label{Viet}
\begin{array}{llll}  
\Sigma x_k = - f_{N-1}(t)/f_N(t),~~\Sigma y_k = - g_{N-1}(t)/g_N(t),~~ \\
\Sigma x_ix_j =f_{N-2}/f_N(t),~~ \Sigma y_iy_j =g_{N-2}/g_N(t),~~ \\
.............................................................................................\\
x_1x_2...x_N=(-1)^N f_0(t)/f_N(t),~~ y_1y_2...y_N=(-1)^N g_0(t)/g_N(t)
\end{array} 
\end{equation}
(summation over all the roots with $k=1,2,..N$,~~ $i,j=1,2,..N, ~~i<j,~ \ldots$, 
is accepted above and throughout the paper).
  
Let us suppose now that the coefficients of the general polynomial
system (\ref{polynom}) {\it are also polynomials with respect to the
  time-like parameter $t$}. If the latter is considered on the same
foot as the coordinates $x,y$, one can easily see that, in the general {\it ``nondegenerate''} case (see below), the coefficients in the higher degree monomials
$a_{n,0},a_{n-1,1},\ldots a_{0,n}$ and $b_{m,0},b_{m-1,1},\ldots b_{0,m}$ {\it
  do not depend on time at all}, the coefficients in the next, lower
order monomials depend on $t$ linearly, and so on. 

Further, during the procedure of elimination, say of $x$, the property
of {\it homogeneity} of corresponding sums of the monomials of the same degree in the eliminant $R_x(y,t)$ with
respect to $y$ and $t$ will be evidently preserved.  This is
equivalent to the statement that {\it the coefficients $g_N$,
  $g_{N-1}$, $g_{N-2}$, \ldots  will be of order $0,1,2,\ldots $ in $t$,
  respectively}.  The same can be certainly said about the
coefficients $f_N\sim t^0,~f_{N-1}\sim t^1,~f_{N-2}\sim t^2,\ldots $ in
$R_y(x)$.

In particular, the coefficients $g_N,f_N$ in the leading terms of the
eliminants depend only on the corresponding coefficients in the
leading monomials ~\cite{Brill,Utyashev} and are, therefore, {\it
  constants} in the case under consideration. The exact connection of
these coefficients had been computed in~\cite{Brill,Utyashev} and,
generally, is of the following form:
\begin{equation}\label{coeffconnect}       
g_N \equiv f_N = {\mathrm{Res}}[F_1^{(n)}(\xi,1)],F_2^{(m)}(\xi,1),\xi]\equiv 
{\mathrm{Res}} [F_1^{(n)}(1,\xi),F_2^{(m)}(1,\xi),\xi], 
\end{equation}
where $F_1^{(n)}(x,y),~F_2^{(m)}(x,y)$ are the sums of the leading monomials, of
degrees $n$ and $m$ respectively, of the generating polynomials in
(\ref{polynom}), and $Res[..]$ designates corresponding resultant of
these two over $\xi$, under the substitution $x=\xi,~y=1$ or,
equivalently, $x=1,~y=\xi$. Below we assume that {\it the principal coefficients (\ref{coeffconnect}) are nonzero}. 
The illustration of the above described construction will be presented in section 4. 

Let us now look once more at the Vieta formulas (\ref{Viet}). The
right-hand sides of formulas of the $I$-th order in roots are proved to be
polynomials of the $I$-th degree in $t$. Therefore, after $I$
differentiations by $t$ of the corresponding formula, one necessarily
arrives at a constant in the right-hand side of the final
equation. This means that {\it there exists a whole set of
  time-independent relations (constraints) between the positions,
  velocities, accelerations, etc. of the particles-roots}!

On the other hand, the {\it canonical} representation of conservation
laws has the form of a sum of various characteristics of {\it
  individual} particles.  In order to obtain such a form from the
above established relations, let us make use of the so-called {\it
  Newton's identities}~\cite{NewtonId} which easily
allow to express the left-hand sides of (\ref{Viet}) as {\it the sums of $I$-th degrees 
of the roots $\{x_k\}$ or $\{y_k\}$},
respectively. As a result, one obtains the following formulas
establishing dynamical correlations between the roots-particles of the
generating system (\ref{polynom}):
\begin{equation} \label{VietM}
\begin{array}{llll}  
\Sigma x_k = A_1(t),~~\Sigma y_k = B_1(t),~~ \\
\Sigma x_k^2 =A_2(t),~~ \Sigma y_k^2 =B_2(t),~~ \\
.........................................\\
\Sigma x_k^N = A_N(t),~~\Sigma y_k^N = B_N(t),
\end{array}
\end{equation}
where $A_k(t)$, $B_k(t)$ are some polynomials in $t$ of degree $k$, respectively.
 
Now again one can differentiate by $t$ each of the {\it modified Vieta
  formulas} (\ref{VietM}) appropriate number of times to obtain a
constant in its right-hand side. In this way one evidently comes to
{\it a chain of conservation relations} which contain the sums of
combinations of the roots, separately of $\{x_k\}$ and $\{y_k\}$, and their higher order time
derivatives and do hold for any generating system of equations
(\ref{polynom}), under the above described, {\it quite general}
polynomial dependence of the coefficients on time $t$.  Thus, {\it any
  system of two nondegenerate polynomial equations in $x,y,t$ defines a
  collective dynamics of $N=nm$ particles-roots for which a set of
  $2(N-1)$ conservation laws does exist}!

Consider now in more detail the first two modified Vieta formulas
which in fact we have already dealt with in the previous
section. Simplest, linear in roots, Vieta formulas in (\ref{VietM})
represent the dynamics of the center of mass of a system of $N$
identical particles-roots, with coordinates
\begin{equation}\label{center}
X(t)=\frac{1}{N}\sum  x_k,\quad  Y(t)=\frac{1}{N}\sum  y_k,    
\end{equation}
and asserts (due to linear dependence of $A_1,~B_1$ on time $t$) that
the center of mass always moves uniformly and rectilinearly, in
full correspondence with the usual Newtonian mechanics. After first
and second differentiation by $t$, one obtains then the law of
conservation of total momentum and the ``weakened'' form of the
Newton's third law; compare this with the example in the previous
section 2.

As for the next, quadratic in roots, formulas in (\ref{VietM}), their
right-hand sides also depend on $t$ quadratically. That's why, to
obtain a corresponding conservation law, one has to differentiate them
twice and to compose then a $SO(2)$-invariant combination of the
resulting equations for $x$ and $y$ parts.  Then, as it was already
demonstrated at the example in section 2, one comes to the
conservation law for the analogue of total mechanical energy:
\begin{eqnarray}\label{energy2}
\Sigma v_k^2 + \Sigma {\vec a}_k {\vec r}_k = const.
\end{eqnarray}

Both terms in (\ref{energy2}) are real-valued, despite the
contribution of complex conjugate pairs of roots. However, imaginary
parts of the latter contribute to the sums; in particular, the first
term is not positively definite, since
$\Re(v_k^2)=\Re(v_k)^2-\Im(v_k)^2$. Thus, the first term corresponds to the (doubled) total
kinetic energy when only contributions from the $C$-particles related
to complex conjugate pairs of roots can be neglected. Generally, one
should take into account another sort of ``kinetic'' energy which is
negative; its physical meaning, as well as of the second term standing
in place of the potential energy, at present is vague.  

In this connection, nonetheless, it should be noted that the conservation law (\ref{energy2}) is also tightly connected with the well-known {\it virial theorem} from the classical mechanics (see, e.g.,~\cite[p.83]{Golstein}). However, the latter corresponds to the null constant in the right-hand side and, besides, holds 
only {\it in average} for finite movements whereas the constraint (\ref{energy2}) is satisfied at any instant. Remarkably, 
for {\it potential} forces {\it homegeneous} with respect to the mutual distances $r_{ij}$ (so that the pairwise potential energies  $U_{ij}\propto r_{ij}^n,~~n\in \mathbb Z$) the second term in (\ref{energy2}) is equal to $-nU$, where $U$ is the total potential energy~\cite[p.86]{Golstein}. Thus, just in the case $n=-2$ the constraint (\ref{energy2}) exactly reproduces the canonical form of the energy conservation law. That is, the interaction of the form $U\propto 1/r^2$ is, though implicitly, distinguished among others in the framework of ``polynomial mechanics''. We recall, however, that, in the construction under consideration, possibility of the decomposition of the resultant forces (=accelerations) into the pairwise (radial) constituents is not yet proved  if possible. 

Finally, let us consider the Vieta formulas (\ref{VietM}) of higher orders in roots. 
The constraints for time derivatives of corresponding order from the higher order sums 
$\sum x_k^3, \sum y_k^3,\ldots,\sum x_k^N,\sum y_k^N$  
will evidently result in the whole set of (non-canonical) conservation laws. However, 
these constraints  {\it do not allow for $SO(2)$-invariant
(generally, $SO(3)$-invariant) combinations} and are, therefore,
coordinate dependent and, most likely, not physically valuable. 
Nonetheless, this problem certainly deserves further investigation.

\section{The Law of Angular Momentum Conservation}
We are now ready to consider the problem of conservation of the {\it
  total angular momentum}, the last of the set of rotation-invariant
canonical conservation laws. In the 2D case there exists only one
component of the angular momentum vector
\begin{equation}\label{angularZ}
 M_z = \sum  M_k, \quad M_k:= x_k(v_y)_k - y_k(v_x)_k.
\end{equation}
Instantaneous velocities $v_x,v_y$ can be expressed through the
coordinates $x,y$ and $t$. Specifically, differentiating by $t$ the
equations (\ref{resultY}, \ref{resultX}) for eliminants $R_y(x,t)$ and
$R_x(y,t)$, one easily finds:
\begin{equation}
\label{veloc}
v_x(t) = -\frac{\partial_t R_y(x,t)}{\partial_x R_y(x,t)},\quad  
v_y(t) = -\frac{\partial_t R_x(y,t)}{\partial_y R_x(y,t)}, 
\end{equation}
so that one obtains the desired expression for angular momentum.
Composing now the auxiliary function
\begin{equation}\label{auxil}
H(M,x,y,t): = M-(xv_y-yv_x), 
\end{equation}
reducing it to common denominator and equating then the numerator to zero, 
one arrives at the additional polynomial equation of the form 
\begin{equation}\label{number}
  M \partial_yR_x(y)\partial_xR_y(x)+x\partial_tR_x(y)\partial_xR_y(x)-y\partial_tR_y(x)\partial_yR_x(y)=0.  
\end{equation}

Together with the generating equations (\ref{polynom}) themselves,
equation (\ref{number}) constitute a complete system of three
polynomial equations for determination of the values of angular
momenta and their temporal dependence of all the particles-roots.  For
this, we must successfully eliminate both $x$ and $y$, again making
use of the resultants' techniques, and arrive at a polynomial equation
of the form $F(M,t)=0$. As a rule, it consists of two multipliers from
which only one corresponds to the true solutions of the system under
consideration whereas the second factor leads to the {\it redundant}
solutions~\footnote{
Appearance of the redundant factor is, probably, the
  consequence of discarding the denominator during the transformation
  of the function (\ref{auxil})}. Separating the proper factor (which
should be exactly of the degree $N=nm$) and making then use of the
linear Vieta formula, one obtains the value of total angular momentum
$M_z$ and verifies that {\it it is always time
  independent}!

Let us now illustrate the above-presented procedure at a typical
example of a system of two independent and {\it nondegenerate} polynomial (with respect to both $x,y$ and $t$) equations of the degrees $n=3$ and $m=2$, 
respectively, with {\it   arbitrary chosen (integer) real-valued coefficients}. Specifically,
let us take
\begin{equation}\label{genexample}
\begin{array}{llll}
F_1=(3x^3+7y^3+6t^3-2x^2y+5xy^2+7t^2x-4t^2y\\
~~~+6tx^2-5ty^2-9xyt)+(-x^2-3y^2-9t^2-xy-10tx-11ty)\\
~~~+(3x+2y-13t)-8=0,\\
F_2=(7x^2-12y^2-4t^2+17xy+5tx-11ty)\\
~~~+(19x+21y+3t)+1=0.
\end{array}
\end{equation}
Let us now rewrite the above system in the following equivalent form. 
Collecting the monomials in $x$ and $y$ and fixing in this way the coefficients in their terms  depending on $t$, one obtains the sought for structure of the generating system (\ref{genexample}):
\begin{equation}\label{gensystem}
\begin{array}{llll}
F_1 =
(3x^3-2x^2y+5xy^2+7y^3)+(6t-1)x^2-(9t+1)xy- \\
~~~~~~~~(5t+3)y^2 +(7t^2-10t+3)x-(4t^2+11t-2)y+ \\
~~~~~~~~(6t^3-9t^2-13t-8)=0,~~ \\
F_2=(7x^2+17xy-12y^2)+(5t+19)x-(11t-21)y- \\
~~~~~~~~(4t^2-3t-1)=0. 
\end{array}
\end{equation}
Thus, we see (cf. with the statement on page 8) that the coefficients in the monomials of 
$(n-I)$-th or $(m-I)$-th degree in $x,y$, respectively, are polynomials in $t$ of the $I$-th degree.

Taking now the resultant of $F_1$ and $F_2$ over $y$ and under
fixed $x$, one obtains the eliminant equation of the form
\begin{equation}\label{elimX}
\begin{array}{lll}
R_y(x,t)=-358343 x^6+(-374447 t-1176858) x^5+ \\   
\left(103876 t^2-864563 t-1155352\right) x^4 +...\\
   +(-111802 t^6-112903 t^5+...+370780)=0,
\end{array}
\end{equation}
and, acting in full analogy, the dual equation
\begin{equation}\label{elimY}
\begin{array}{lll}
R_x(y,t)=-358343 y^6+(145966 t-95313) y^5+\\
  \left(2828 t^2+968047 t-354945\right) y^4+...\\
  +(4148 t^6-40308 t^5+...+222804)=0.
\end{array}
\end{equation}

Again, the coefficients in the terms of $(N-I)$-th degree (with $N=6$) in the eliminants (\ref{elimX},\ref{elimY}) are polynomials in $t$ of the $I$-th degree (cf., once more, with the  corresponding statement on page 8). In particular, both leading coefficients in the eliminants $f_6=g_6=-358343$ are equal and can be equivalently computed via formula (\ref{coeffconnect}). They do not turn to zero so that the system (\ref{genexample}) is indeed nondegenerate.

\noindent\hspace{4ex}Thus, we can presuppose that this system, of a quite general~form, uniquely
defines the correlated dynamics of $N=3\times 2 =6$ particles of equal masses (some of which being merged into complex conjugate pairs and possessing, therefore, a twice greater mass) for which {\it the laws of conservation of total momentum, total angular momentum   and the analogue of total mechanical energy are valid all three together}!

At any $t$, equations (\ref{elimX}) and (\ref{elimY}) both have 6
solutions some of them being real and some entering in complex
conjugate pairs. Putting these in correspondence to each other, one
obtains all 6 solutions of the generating system
(\ref{gensystem}). Further, the {\it discriminants} of the eliminants
(\ref{elimX}) and (\ref{elimY}) contain a common factor $D(t)$ which
is a polynomial in $t$ of the order 18 and turns to zero at 18 values
of $t$; however, only 2 of them are real. Thus, there are precisely 2 
instants at which some two of the roots of (\ref{gensystem})
become {\it multiple}; physically, these correspond to the ``events''  of
annihilation of a pair of $R$-particles or, conversely, creation of a
pair from a composite $C$-particle.

Making use of the linear Vieta formulas for (\ref{elimX}) and
(\ref{elimY}), one immediately finds that the center of mass of the
``closed mechanical system'' of 6 particles moves uniformly with the
(dimensionless) velocity ${\vec V}=\{V_x,V_y\},~~V_x=-374447/358343,~V_y=145966/358343$.

From the second order (modified) Vieta formulas for $x$ and $y$ parts
one finds after two differentiations by $t$ and summation of both:
\begin{equation}\label{energy3}
(v_1^2+\ldots +v_6^2)+(\vec a_1\vec r_1+\ldots +\vec a_6\vec r_6) = const =
\frac{237989891909}{358343^2}\approx 1.85336.
\end{equation}

Calculating now the expression for numerator of the defining function
of angular momentum (\ref{auxil}), one obtains the (rather
complicated) polynomial equation of the form (\ref{number}),
\begin{equation}\label{complicang}
  \Phi(M,x,y,t)=0.
\end{equation}

Eliminating successively $x$ and $y$ from the joint system of
equations (\ref{gensystem}) and (\ref{complicang}), via calculating
corresponding resultants, one arrives at the polynomial equation for angular
momenta of the particles that turns out to be of the form
$F(M,t)G(M,t)=0$, with $F$ and $G$ being polynomials in $M$ of degrees
90 (!)  and 6, respectively. So it is evident (and can be verified
through numerical calculations of the roots and related angular
momenta) that the first factor leads to redundant solutions whereas
the equation
\begin{equation}\label{angmom3}
G(M,t)=\alpha(t) M^6+\beta(t) M^5 + \gamma(t) M^4+\ldots +\varepsilon(t) =0
\end{equation}
properly defines the values of angular momenta of all
6 particles at any instant of time!
In (\ref{angmom3}) the coefficients $\alpha(t)=358343A(t)$ and $\beta(t)=827188A(t)$ {\it turn out to be  proportional to each other} by the factor $A(t)=A_0 t^{18}+A_1 t^{17}+...+A_{18}$ which is a polynomial
~\footnote{Remarkably, this polynomial is always proportional to the above-mentioned common factor $D(t)$ of the 
discriminants that defines the instants of ``events''}  of 18-th degree in $t$ with very large integer numerical coefficients $\{A_j\},~j=1,..18$. Now, making use of the linear Vieta formula for the equation (\ref{angmom3}), one immediately obtains the
value of {\it total} angular momentum of 6 particles:
\begin{equation}\label{angtotal}
  M_z = \sum  M_k = -\frac{\beta(t)}{\alpha(t)} = const = -\frac{827188}{358343}\approx -2.308
\end{equation}
which, remarkably,  does not depend on $t$! 
Of course, this result had been also reproduced via direct numerical
calculations under numerous values of the time parameter $t$. Considerable 
number of other examples of the algebraic systems of equations with nondegenerate 
polynomials in $x,y$ and $t$, of different and rather high degrees (say, $n=7,~m=5$) had been studied; for all the examples three canonical-like conservation laws 
are undoubtedly satisfied, and the corresponding physical characteristics 
were exactly computed.  
 
\section{Conclusion}

In the paper we have demonstrated that for a very wide class of
systems of polynomial equations, with polynomial dependence of the
(real-valued) coefficients on the time parameter, there exists a whole
set of {\it time-independent} constraints on the roots of the system
and their time derivatives. Real roots can be identified with one sort
({\it R-}) of identical particlelike formations while complex
conjugate with the other one, {\it C}-particles, possessing a twice
greater mass and participating in the processes of annihilation or
creation of a pair of {\it R}-particles.

Thus, one gets a nontrivial correlated (at present, 2D
nonrelativistic) dynamics of the ensemble of {\it R-} and {\it
  C-}particles. This can model real physical dynamics in a system of
interacting particles and even replace its canonical description on
the base of the Newton's differential equations of motion.

Time-independent constraints arising in our scheme are generated by
the set of the Vieta formulas that impose rigid restrictions on the
instantaneous positions of particles-roots, their velocities,
accelerations, etc. We have shown that, generally, these constraints
can be transformed into the form of conservation laws. Moreover, all
the canonical rotation-invariant conservation laws are represented
herein. In particular, for any nondegenerate polynomial (with respect both to $x,y$
and $t$) system of equations the laws of conservation
of total momentum, angular momentum and (the analogue of) total
mechanical energy are satisfied!

In the framework of the concept of the ``unique worldline'', there are
a lot of both mathematical and physical problems to be solved. As for
mathematics, one has every reason to think that a sort of {\it
  generalized Vieta-like formulas} relating different roots of any
system of {\it two or more polynomial equations} do exist and could be
discovered; this is an interesting challenge for ``pure''
mathematicians.  In particular, such formulas, in contrast with the familiar 
1D case, could explicitly {\it mix} different coordinates $x,y,\ldots $ of
the solutions and, under the presence of the time-like parameter,
their time derivatives. These conjectural relations would help to
explain, say, the effect of angular momentum conservation which at
present is confirmed only on an essential number of examples but not
in a general analytical form~\footnote{For particular classes of
  polynomial systems reliable ``phenomenological'' formulas for
  time-independent values of total angular momentum have been fitted;
  these will be presented elsewhere}.

As for physics (which is also completely induced herein by the purely
mathematical properties of polynomials), one should generalize the
scheme to the physical 3D case and find the road to the relativistic
reformulation of the theory. Apart of these obvious goals, the problem
of determination of an effective pairwise particle interaction' force
is on the agenda. And, of course, one should analyze the meaning of
the term which stands for the ordinary potential energy in the
corresponding conservation law (\ref{energy2}). The principal
question is, however, the following: is it possible to form composite
and stable multi-particle {\it clusters} modelling the real elementary
particles, nuclei, etc.  from the initial {\it R- and C- pre-elements
  of matter} naturally arising in our scheme? And how could be then
interpreted the latter from the physical viewpoint?

In general, there are a lot of remarkable relations in mathematics,
and in the {\it nonlinear algebra} in particular~\cite{Morozov2},
which are not yet discovered but could be responsible for the
structure of fundamental physical laws. Here we have undertaken one
more attempt~\footnote{For a related ``algebrodynamical'' approach,
  see, e.g., our works~\cite{AD,YadPhys} and references therein} to
shed some light upon the deep connections existing between fundamental
physics and mathematics.

\section*{Acknowledgement}
 \thanks{The authors are grateful to A.V. Koganov, M.D. Malykh, G. Nilbart, 
J.A. Rizcallah and especially to A. Wipf for friendly support and valuable discussions on the subject.}

\bibliography{ArxivVestnikKass}{}
\bibliographystyle{unsrt}

\end{document}